# Tracking Walls, Take-It-Or-Leave-It Choices, the GDPR, and the ePrivacy Regulation

**Frederik J Zuiderveen Borgesius, Sanne Kruikemeier, Sophie C Boerman and Natali Helberger***

* Dr Frederik J Zuiderveen Borgesius is a researcher at Research Group on Law Science Technology & Society (LSTS) of the Free University Brussels. His email address is: fzuiderv@vub.ac.be. Dr Sanne Kruikemeier is assistant professor Political Communication and Journalism at the Amsterdam School of Communication Research (ASCoR) of the University of Amsterdam. Dr Sophie C Boerman is assistant professor of Persuasive Communication at the Amsterdam School of Communication Research (ASCoR) of the University of Amsterdam. Prof Dr Natatali Helberger is professor Information Law at the Institute for Information Law (IViR) of the University of Amsterdam.[1]

[1] All authors cooperate in the Personalised Communication Project <http://personalised-communication.net>. The research for this article was made possible with the help of research funding from the University of Amsterdam (Personalised Communication - PI's Prof Dr Claes de Vreese and Prof Dr Natali Helberger), and a research grant from the European Research Council, ERC, under Grant 638514 (PersoNews - PI N Helberger). The authors would like to thank Joyce Neys, Candida Leone,

*Abstract*

*On the internet, we encounter take-it-or-leave-it choices regarding our privacy on a daily basis. In Europe, online tracking for targeted advertising generally requires the internet users' consent to be lawful. Some websites use a tracking wall, a barrier that visitors can only pass if they consent to tracking by third parties. When confronted with such a tracking wall, many people click 'I agree' to tracking. A survey that we conducted shows that most people find tracking walls unfair and unacceptable. We analyse under which conditions the ePrivacy Directive and the General Data Protection Regulation allow tracking walls. We provide a list of circumstances to assess when a tracking wall makes consent invalid. We also explore how the EU lawmaker could regulate tracking walls, for instance in the ePrivacy Regulation. It should be seriously considered to ban tracking walls, at least in certain circumstances.*

# I. Introduction

On the internet, we encounter many take-it-or-leave-it choices regarding our privacy. Social network sites and email services typically require users to agree to a privacy statement or to terms and conditions – if people do not agree, they cannot use the service. Some websites use a tracking wall, a barrier that visitors can only pass if they agree to tracking by third parties. When confronted with such take-it-or-leave-it choices, many people might click 'I agree' to any request. It is debatable whether people have meaningful control over personal information if they have to consent to tracking to be able to use services or websites. This article explores the problem of tracking walls, and discusses several options for the EU lawmaker to regulate them.

First, we give a brief introduction to online tracking and tracking walls. Next, we report on a large-scale survey we conducted in the Netherlands: most people consider tracking walls unfair, and do not find it acceptable when they have to disclose their personal data in exchange for using a website.

Then we analyse whether such tracking walls are allowed under European data privacy law, more specifically the ePrivacy Directive, and the General Data Protection Regulation (GDPR).[2] The European Commission has published a proposal for an ePrivacy Regulation, which should replace the ePrivacy Directive.[3] That ePrivacy proposal does not contain specific rules regarding tracking walls. We explore how the EU lawmaker could deal with tracking walls, and discuss four options: (i) no specific rules for tracking walls; (ii) ban tracking walls in certain circumstances; (iii) fully ban tracking walls; (iv) ban all web-wide tracking (third party tracking). We argue that the lawmaker should seriously consider options (ii) and (iii): a partial or complete ban of tracking walls.[4]

## II. Online Tracking

People can use many online services, such as websites, email services, and search engines, without paying money. Large amounts of data about users are usually collected through such services. Those data are typically used for behavioural targeting, a marketing technique which involves monitoring people's online behaviour, and using the collected information to show people individually targeted advertising.[5]

In a simplified example, behavioural targeting works as follows. Advertising networks are companies that serve advertising on websites. Such ad networks track people's browsing behaviour over multiple websites, using cookies or similar technologies. (We speak of 'cookies' for readability, but ad networks use many other tracking technologies too, such as flash cookies and device fingerprinting.[6]) By following people's browsing behaviour, ad networks build a profile of people's inferred interests, to show them more targeted ads. For instance, an ad network may show wine advertising to people who visited websites about wine. Many websites allow dozens of ad networks and other third parties to store tracking cookies on their visitors' computers.[7] Tracking people across multiple websites (for instance by ad networks) is called 'web-wide tracking' or 'third party tracking'. Tracking within one website could be called 'site-wide tracking'.[8]

---

[2] We use the phrase 'data privacy law' if we refer to both the rules in the ePrivacy Directive and in data protection law.

[3] Proposal for a Regulation of the European Parliament and of the Council, concerning the respect for private life and the protection of personal data in electronic communications and repealing Directive 2002/58/EC (Regulation on Privacy and Electronic Communications) COM(2017) 10 final, <https://ec.europa.eu/digital-single-market/en/news/proposal-regulation-privacy-and-electronic-communications> accessed 7 September 2017.

[4] The paper builds on, and includes sentences from: Frederik J Zuiderveen Borgesius, *Improving privacy Protection in the Area of Behavioural Targeting* (Kluwer law International 2015) and Frederik J Zuiderveen Borgesius et al, 'An Assessment of the Commission's Proposal on Privacy and Electronic Communications' (Directorate-General for Internal Policies, Policy Department C: Citizen's Rights and Constitutional Affairs, May 2017) https://ssrn.com/abstract=2982290 accessed 7 September 2017.

[5] Sophie Boerman, Sanne Kruikemeier and Frederik J Zuiderveen Borgesius, 'Online behavioral advertising: a literature review and research agenda' (2017) 3 Journal of Advertising 363-376.

Websites are often funded by advertising. Sometimes, advertisers only pay a website publisher if somebody clicks on an ad. Few people click on ads. Roughly, if an ad is shown 10,000 times, less than five people click on that ad (a 'click through rate' of 0.05%).[9] Behavioural targeting was developed to make more people click on ads. Some claim that behavioural targeting is more effective than non-targeted ads, and that it leads to more income for website publishers.[10]

However, in the long term, behavioural targeting may lead to less income for certain website publishers. With traditional advertising models, advertisers had to advertise in certain media to reach certain people. The price of an ad is based, among other things, on the number of readers.[11] By way of illustration, a printed newspaper with many wine lovers among its readers could be a good place for a wine merchant to advertise. And the book review pages of a printed newspaper could be a good place to advertise a book. The newspaper assembles an audience, and provides the advertiser access to this audience. Contextual advertising is a type of online advertising that resembles print advertising, as ads are placed in the context of related content. With contextual online advertising, advertisers aim to reach people by showing ads on certain websites or pages – for example ads for wine on websites about wine.

---

[6] Chris Jay Hoofnagle et al, 'Behavioral advertising: the offer you cannot refuse' (2012) 6 Harvard Law & Policy Review 273.

[7] Ibrahim Altaweel, Nathaniel Good and Chris Jay Hoofnagle, 'Web Privacy Census' (*Technology Science*, 15 December 2015) http://techscience.org/a/2015121502 accessed 7 September 2017.

[8] World Wide Web Consortium, 'Tracking Preference Expression (DNT), W3C Candidate Recommendation 20 August 2015' (2015) <https://www.w3.org/TR/tracking-dnt> accessed 7 September 2017; Article 29 Working Party, 'Opinion 01/2017 on the Proposed Regulation for the ePrivacy Regulation (2002/58/EC)' (4 April 2017) 18.

[9] Dave Chaffey, 'US, Europe and Worldwide display ad clickthrough rates statistics summary' (*Smartinsights*, 8 March 2017) <http://www.smartinsights.com/internet-advertising/internet-advertising-analytics/display-advertising-clickthrough-rates/> accessed 7 September 2017: 'Across all ad formats and placements Ad CTR is just 0.05%. So, this is just than 5 clicks per 10000 impressions (...)'

[10] See, Howard Beales, 'The value of behavioral targeting' (Network Advertising Initiative, 2010) < https://www.networkadvertising.org/pdfs/Beales_NAI_Study.pdf > accessed 31 August 2017. See also, Katherine J Strandburg, 'Free Fall: The Online Market's Consumer Preference Disconnect' (University of Chicago Legal Forum, 2013) 95 http://chicagounbound.uchicago.edu/cgi/viewcontent.cgi?article=1511&context=uclf accessed 28 September 2017.

By contrast, with behavioural targeting, an ad network can show a wine ad anywhere on the web to people whose profile suggests that they like wine. An ad network does not have to buy expensive ad space on a large professional website, such as a newspaper website, to advertise to an individual. The ad network can reach that individual when he/she visits a small website, where advertising space is cheaper.[12] Hence, the ad network spends less money on large professional websites. In sum, advertising funds many services. But in the long term, behaviourally targeted advertising may reduce income for certain website publishers.

## III. Tracking Walls

It is widely assumed that online tracking and behavioural targeting raise serious privacy concerns.[13] In Europe, online tracking for targeted advertising generally requires the internet users' consent to be lawful (see Section V below). Companies use different strategies to collect people's consent to online tracking. One strategy is offering people a take-it-or-leave-it-choice. For example, some websites use 'tracking walls,' also called 'cookie walls' – barriers that visitors can only pass if they allow the website or its partners to track them.

---

[11] Fernando Bermejo, *The internet audience: Constitution & measurement* (Peter Lang 2007).
[12] ibid; see also, Joseph Turow, *The Daily You: How the New Advertising Industry is Defining Your Identity and Your Worth* (Yale University Press 2011); Alexis Madrigal, 'A Day in the Life of a Digital Editor, 2013' *The Atlantic* (6 March 2013) < https://www.theatlantic.com/technology/archive/2013/03/a-day-in-the-life-of-a-digital-editor-2013/273763/> accessed 31 August 2017.
[13] See for instance, Turow (n 11); Julia Angwin, *Dragnet Nation: A Quest for Privacy, Security, and Freedom in a World of Relentless Surveillance* (Times Books 2014); Article 29 Working Party, 'Opinion 2/2010 on online behavioural advertising' (22 June 2010) WP 171.

Below we give an example of such a tracking wall.[14] People can only enter the website if they click 'I agree' and thus consent to web-wide tracking. If people do not click 'I agree,' they cannot enter the website.

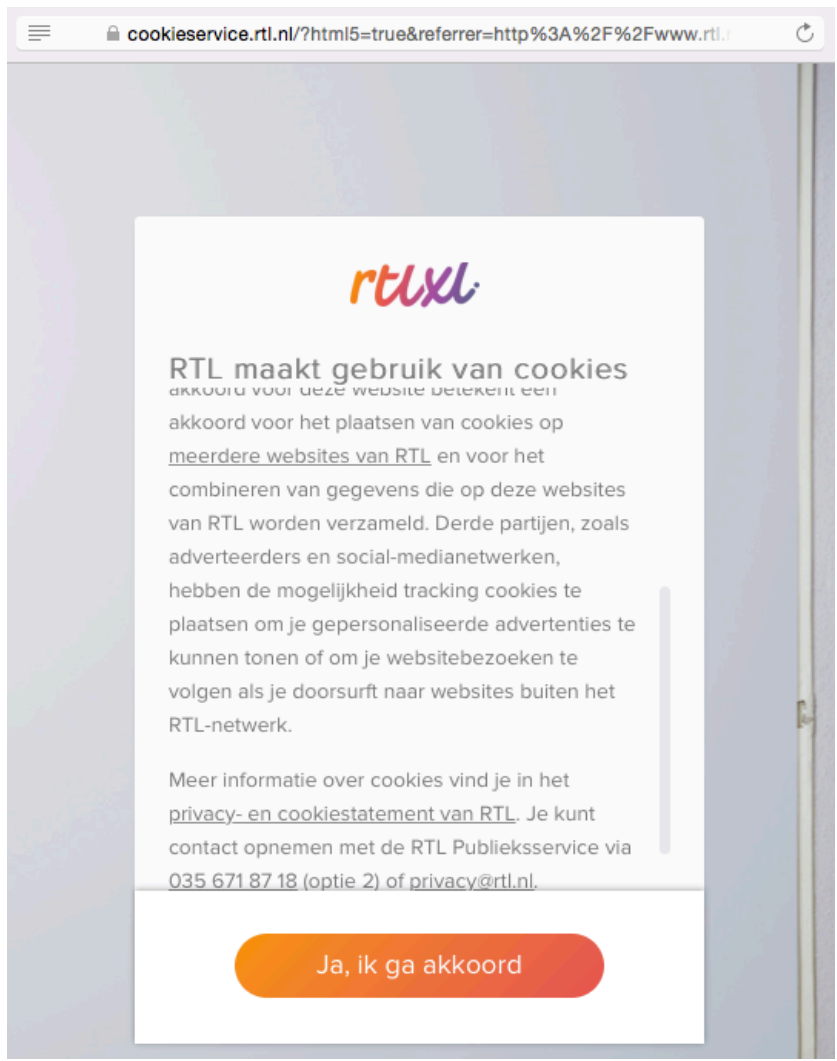

Figure 1. An example of a tracking wall.

---

[14] See for another example, Ronald Leenes and Eleni Kosta, 'Taming the cookie monster with Dutch law – A tale of regulatory failure' (June 2015) 31(3) Computer Law & Security Review 317-335.

Text in the pop-up (informal translation):

'RTL uses cookies

By clicking on "Yes, I agree" you give RTL permission to collect, with cookies and similar technologies, (personal) data, to store these data, and to process these data as described in our privacy and cookie statement. These data, for instance about your location and about which videos you watch, are used to improve our service and to show you advertisements and videos which are tailored to your use. This website is part of the RTL network. Consent on this website implies consent for placing cookies on multiple websites or RTL, and for combining data that are collected on these websites of RTL. Third parties, such as advertisers and social media networks, can place tracking cookies to show you personalised advertisements or to follow your website visits if you surf to websites outside the RTL network.

You can find more information about cookies in the privacy and cookie statement of RTL. You can contact the RTL customer service on 035 671 87 18 (option 2) or privacy@rtl.nl.

Yes, I agree.'[15]

## IV. Survey Results: Most People Find Tracking Walls Unacceptable

We conducted a survey among the Dutch public, to examine the public opinion regarding tracking walls. A total of 1,235 people answered the survey questions. This sample is representative for the Dutch population: the findings can be extrapolated to the general public.[16] We introduced our questions about tracking walls with the text:

---

[15] Screenshot 1 March 2017, when visiting <www.rtl.nl> from an IP address in the Netherlands.

[16] In our sample, the average age is 54.1 years old (ranged between 18 and 89 years old) and 51.7% is female. 32.2% of the respondents had a lower level of education, 33.2% has a middle level of education, and 34.4% had a higher level of education (no information about education for two cases). The data were collected in April 2016 by CenterData, a research company managing a research panel <www.centerdata.nl/en>.

'To view or use a [web shop / news website / health website / website with public transportation timetables], you are sometimes required to accept that your personal information is collected through the website. If you do not click 'OK' when encountering a pop-up or a tracking wall, you cannot view the website.' This introductory text was then followed by two questions: 'To what extent do you think this is acceptable?,' and 'To what extent do you think this is fair?'[17] The results are presented in Figures 2 and 3.

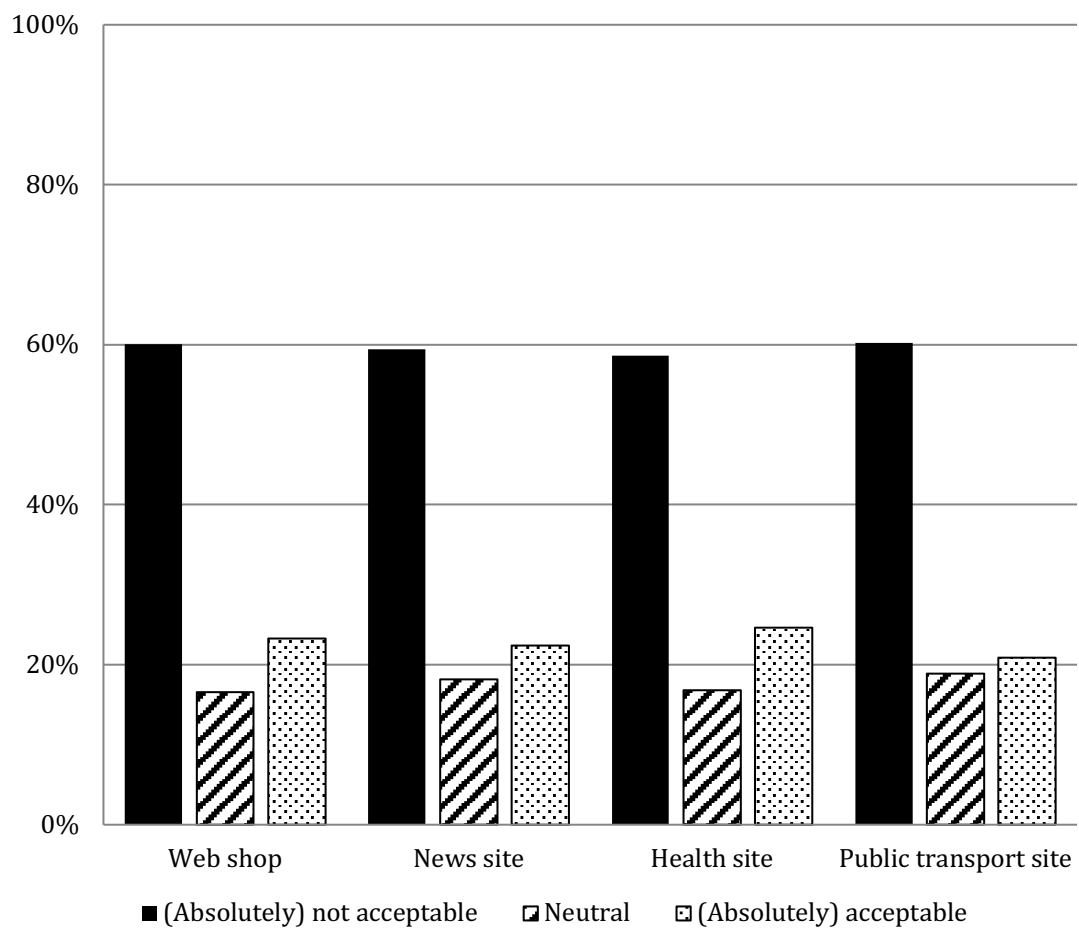

*Figure 2. The percentage of respondents who think tracking walls are (not) acceptable (N = 1235).*

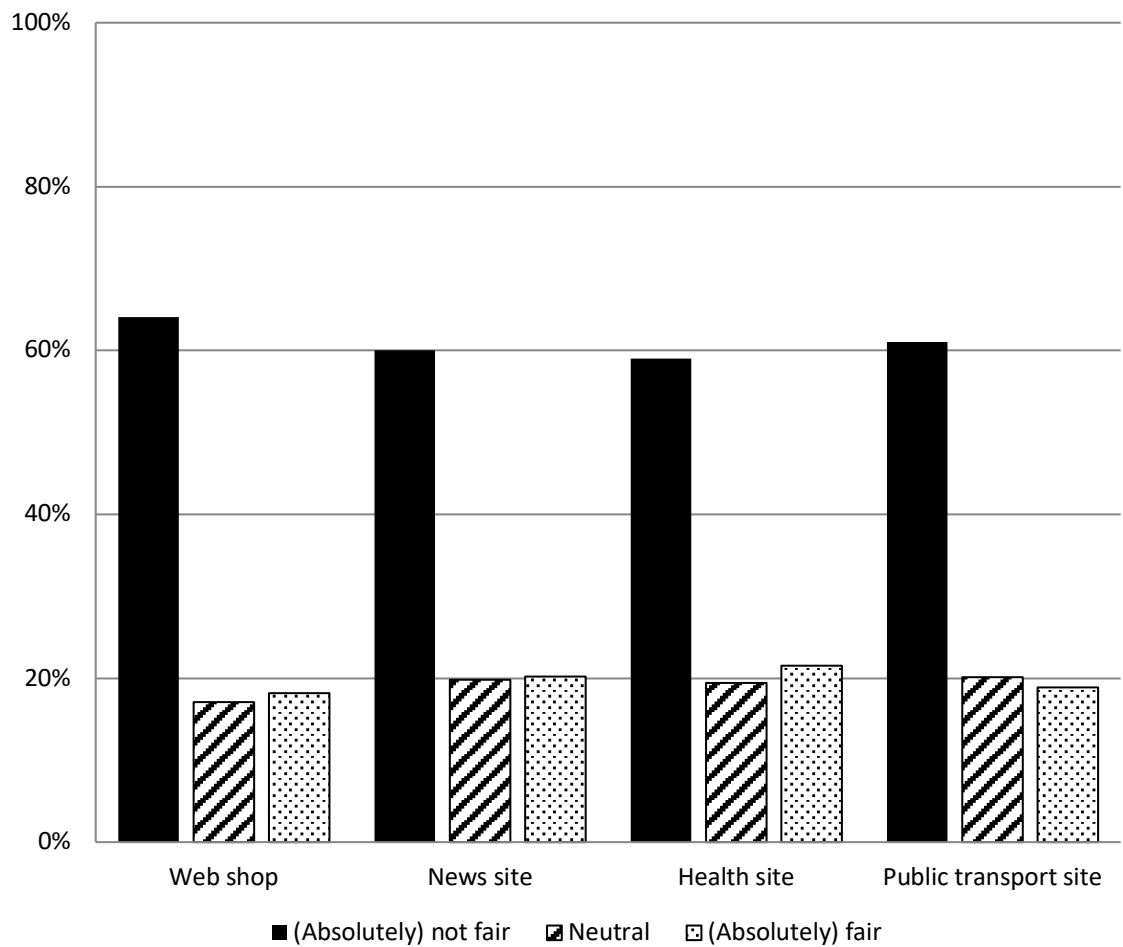

*Figure 3. The percentage of respondents who think tracking walls are (not) fair (N = 1235).*

The results show that a majority of people (roughly 60%) believes it is neither acceptable nor fair when websites use tracking walls. Roughly, 20% finds tracking walls acceptable and fair; 20% is neutral. There do not appear to be substantial differences between the types of website – shopping, news, health, or public transportation. For all these website types, most people believe tracking walls are unacceptable and unfair.

People are used to trade-offs when using media or websites. Making access to 'free' content conditional upon the acceptance of advertising, for example, has a long tradition in the media sector, as advertising funds many media. Do people find certain trade-offs more acceptable or fair than others?

We asked people two questions about trade-offs: a trade-off where they have to disclose personal data in exchange for using a website, and a trade-off where they have to accept that there are ads on a website. The first question started with the following text: 'Websites are offered for free. This is possible because the website lets other companies collect personal information through the website, for instance about the things you click on.' This introductory text was then followed by the following questions: 'To what extent do you think it is acceptable if websites are free, in exchange for collecting your personal information?'

The next question started with the following text: 'Websites are offered for free. This is possible because the website contains ads.' This introductory text was then followed by the following question: 'To what extent do you think it is acceptable if websites are free, in exchange for showing ads?'[18] These questions thus provide insights into how acceptable people find it if, in exchange for visiting a so-called 'free' website, they have to accept that their data are collected, or that a site includes advertising. The results are presented in Figure 4.

---

[18] Respondents could answer these questions on a seven-point scale ranging from 1 (Absolutely not) to 7 (Absolutely). We recoded these answers for presentation in the Figures: answers 1 till 3 were recoded as *not acceptable*; answer 4 as *neutral*; answers 5 till 7 as *acceptable*.

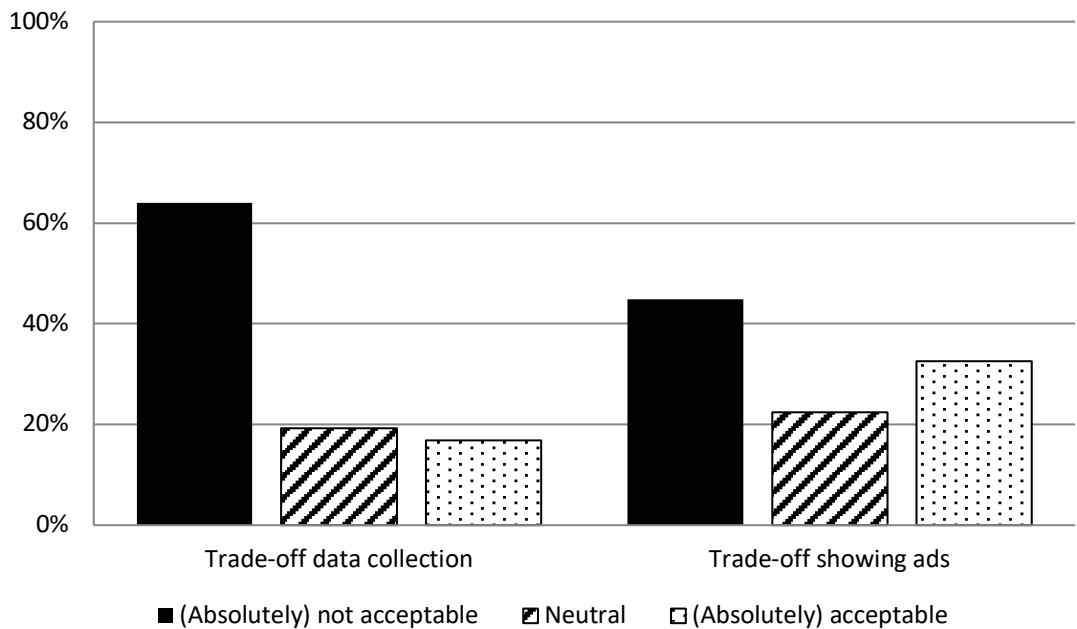

*Figure 4. The percentage of respondents who believe (i) the collection of personal data or (ii) advertising on websites is (not) acceptable (N = 1235).*

A majority (64.0%) says that trading personal data against the use of a 'free' website is unacceptable. Only 44.9% thinks that is not acceptable if a 'free' website contains ads. Hence, more people seem to dislike a model in which they have to disclose personal data, than a model in which they are exposed to ads.

Nowadays, most online advertising is combined with cookies or other tracking technologies. In practice, people can thus only accept online advertising by also accepting online tracking.[19] However, advertising does not necessarily require tracking and targeting. For instance, tracking people is not necessary for contextual advertising, such as ads for cars on websites about cars.

---

[19] One of the only ways to protect oneself against online tracking is using an ad blocker. See, Craig Wills and Doruk Uzunoglu, 'What Ad Blockers Are (and Are Not) Doing' (Computer Science Department, Worcester Polytechnic Institute, 2016) <http://web.cs.wpi.edu/~cew/papers/tr1602.pdf> accessed 7 September 2017.

Overall, 32.6% of the people accept advertising if they use a 'free' website, and 22.4% are neutral on this issue. The fact that many people accept that 'free' content is advertising-funded, or are indifferent about it, is not surprising.[20] Advertising-funded content has been around for more than a century. Without paying with money, people can listen to commercial radio and read certain advertising-funded newspapers.[21] Trade-offs where companies capture data about their visitors are more recent than trade-offs where people are confronted with advertising.[22]

Our survey results are line with a EU-wide Eurobarometer survey, which found in 2017 that 64% of the respondents finds it unacceptable to 'hav[e] your online activities monitored (for example what you read, the websites you visit) in exchange for unrestricted access to a certain website.'[23]

While most people find tracking walls unacceptable, many people might click 'OK' when being confronted with tracking walls.[24] It seems that many people see disclosing their personal data as inevitable. In an EU-wide survey, 58% agree that '[t]here is no alternative than to provide personal information if you want to obtain products or services.'[25] And 43% agree with the statement 'You feel you have to provide personal information online'.[26] Similarly, surveys in the US show that 'a majority of Americans are resigned to giving up their data'.[27]

---

[20] Our survey results in the Netherlands resemble results from other research. For instance, after interviews in the US in 2010, McDonald and Cranor conclude: 'people understand ads support free content, but do not believe data are part of the deal.' Aleecia McDonald and Lorrie Faith Cranor, 'Beliefs and Behaviors: Internet Users' Understanding of Behavioral Advertising' (38th Research Conference on Communication, Information and Internet Policy, 2 October 2010) 21 <http://ssrn.com/abstract=1989092> accessed 7 September 2017.

[21] See generally: Tim Wu, *The Attention Merchants: The Epic Scramble to Get Inside Our Heads* (Knopf 2016).

[22] It is not new that companies capture data about customers. For instance, shops have used loyalty cards for a long time. Such cards often enable shops to track the shopping behaviour of customers.

[23] European Commission, 'Flash Eurobarometer 443, ePrivacy, full report' (December 2016) Question 5.1 (93/T.24) <https://ec.europa.eu/digital-single-market/en/news/eurobarometer-eprivacy> accessed 7 September 2017.

[24] For instance, in a 2014 survey in the Netherlands, half of the respondents said they usually clicked 'OK' to cookie pop-ups. Dutch Consumer Organisation (*Consumentenbond*), 'Cookiewet heeft bar weinig opgeleverd [Cookie law didn't help much]' (2014) <https://www.consumentenbond.nl/internet-privacy/cookiewet-heeft-weinig-opgeleverd> accessed 7 September 2017.

[25] European Commission, 'Special Eurobarometer 431, Data Protection' (Report, 2015) 29 <http://ec.europa.eu/public_opinion/archives/ebs/ebs_431_sum_en.pdf> accessed 7 September 2017.

There appears to be a 'privacy paradox': in surveys, people say they care about privacy, but people often divulge personal data in exchange for minimal benefits or convenience, and relatively few people use technical tools to protect their privacy online.[28] Scholars from various disciplines argue that people do care about privacy, but have difficulties acting according to their privacy preferences.[29] Similarly, people who care about the environment might not study the label of every supermarket product to establish if it was produced in an environmentally friendly way.

Insights from behavioural studies can help to explain the privacy paradox. 'Present bias', sometimes called myopia, refers to people's tendency to focus more on the present than on the future. People often choose for immediate gratification, and often ignore future costs. many people find it hard to stick with a healthy diet, or to save money for later.[30] Because of present bias, people might click 'I agree' to online tracking if that gives them immediate access to a website – even if they were planning to protect their privacy.[31] If a friend emails a link to a good article, the enjoyment of reading that article is immediate (even if one must accept tracking cookies), while the privacy risks are often further in the future.

---

[26] ibid 43.

[27] Joseph Turow, Michael Hennessy and Nora Draper, 'The Tradeoff Fallacy. How Marketers Are Misrepresenting American Consumers and Opening Them Up to Exploitation' (A report from the Annenberg School for Communication, University of Pennsylvania, June 2015) 1 <https://www.asc.upenn.edu/sites/default/files/TradeoffFallacy_1.pdf> accessed 7 September 2017. See for similar results (based on focus groups), Katharine Sarikakis and Lisa Winter, 'Social Media Users' Legal Consciousness About Privacy' (2017) 3(1) Social Media Society 1-14.

[28] See for instance: Susan Barnes, 'A privacy paradox: Social networking in the United States' (2006) 11(9) First Monday <http://firstmonday.org/ojs/index.php/fm/article/view/1394/1312> accessed 7 September 2017.

[29] See for instance, Sabine Trepte et al, 'What do people know about privacy and data protection? Towards the "Online Privacy Literacy Scale" (OPLIS)' (on file with author, forthcoming) in Gutwirth et al (eds), *Computers, Privacy and Data Protection – Reforming data protection: the global perspective* (Springer forthcoming); Alessandro Acquisti and Jens Grossklags, 'What Can Behavioral Economics Teach Us About Privacy?' in Alessandro Acquisti et al (eds), *Digital Privacy: Theory, Technologies and Practices* (Auerbach Publications, Taylor and Francis Group 2007); McDonald and Cranor (n 19).

[30] Hanneke Luth, 'Behavioural Economics in Consumer Policy: The Economic Analysis of Standard Terms in Consumer Contracts Revisited' (PhD thesis, University of Rotterdam 2010) s 3.2.2, e; Cass Sunstein and Richard Thaler, *Nudge: Improving Decisions about Health, Wealth, and Happiness* (Yale University Press 2008) ch 6.

[31] Alessandro Acquisti, 'Privacy in Electronic Commerce and the Economics of Immediate Gratification' (Proceedings of the 5th ACM Conference on Electronic Commerce,, Association for Computing Machinery, New York 2004) 21–29; Katherine Strandburg, 'Free Fall: the Online Market's Consumer Preference Disconnect' (University of Chicago Legal Forum, 2013) 95, 149 <http://ssrn.com/abstract=2323961> accessed 7 September 2017.

In conclusion, most people find tracking walls unacceptable and unfair, and think it is not acceptable when they have to disclose their personal data in exchange for using a website. In the next section, we discuss how EU data privacy law deals with tracking walls and similar take-it-or-leave-it choices.

## V.  Tracking Walls and EU Data Privacy Law

### 1. The ePrivacy Directive and Tracking Walls

In European data privacy law, informed consent and individual choice play a central role, like in many data privacy rules around the world.[32] The ePrivacy Directive requires, in short, companies to ask prior the internet user's prior consent before they use tracking cookies and similar technologies.[33] The Directive contains exceptions to the consent requirement, for cookies that are necessary for communication and for services requested by the user.[34] The Directive's consent requirement applies to many tracking technologies, such as several types of cookies, and some forms of device fingerprinting.[35] For the definition of consent, the ePrivacy Directive refers to the general data protection law (the Data Protection Directive).[36] Under the Data Protection

---

[32] See on data privacy law around the world, Graham Greenleaf, 'Global Data Privacy Laws 2017: 120 National Data Privacy Laws, Including Indonesia and Turkey' (30 January 2017) 145 Privacy Laws & Business International Report, 10-13 <https://ssrn.com/abstract=2993035> accessed 7 September 2017.
[33] art 5(3) ePrivacy Directive 2002/58/EC (updated in 2009). General data protection law probably also requires the internet user's consent for most online tracking practices. See, Frederik J Zuiderveen Borgesius, 'Personal Data Processing for Behavioural Targeting: Which Legal Basis?' (2015) 5(3) International Data Privacy Law 163–176 doi: 10.1093/idpl/ipv011. But in this paper, we focus on the consent requirement in the ePrivacy Directive.
[34] art 5(3) ePrivacy Directive 2002/58/EC (updated in 2009). See, Article 29 Working Party, 'Opinion 04/2012 on Cookie Consent Exemption' (7 June 2012) WP 194.
[35] Eleni Kosta, 'Peeking into the cookie jar: the European approach towards the regulation of cookies' (2013) 21 International Journal of Law and Information Technology 1.
[36] For the definition of consent, art 2(f) ePrivacy Directive 2002/58/EC (updated in 2009) refers to the Data Protection Directive.

Directive, valid consent requires a (i) freely given, (ii) specific, and (iii) informed (iv) indication of wishes.[37]

Hence, consent must be 'freely given' to be valid. For instance, say an employer asks an employee for consent to process his or her personal data. Such consent might not be 'freely given', because of the power imbalance. Employees might fear adverse consequences if they do not consent.[38] The Court of Justice of the European Union says that people applying for passports cannot be deemed to have freely consented to have their fingerprints taken, because people need a passport.[39] In short, consent is not 'freely given', and therefore invalid, if people lack a real choice.

From a data protection perspective, the question is whether consent is 'freely given' when people agree to web-wide tracking after encountering a tracking wall.[40] Among others, Kosta suggests that a tracking wall makes consent involuntary: 'In such a case the user does not have a real choice, thus the consent is not freely given.'[41] Indeed, in some cases consent may not be 'freely given' when a website uses a tracking wall. For example, the Dutch Data Protection Authority says that the national public broadcasting organisation is not allowed to use a tracking wall. The Data Protection Authority says that the public broadcaster has a 'situational monopoly,' because the only way to access certain information online is through the broadcaster's website. This situational monopoly makes people's consent involuntary.[42]

---

[37] art 2(h) Data Protection Directive.
[38] Eleni Kosta, 'Consent in European Data Protection Law' (PhD thesis, University of Leuven 2013) (Martinus Nijhoff Publishers 2013) 386; Article 29 Working Party, 'Opinion 15/2011 on the definition of consent' (13 July 2011) WP 187, 13-14.
[39] Case C-291/12 *Schwartz v Stadt Bochum* [2013] ECLI:EU:C:2013:670, para 32: 'citizens of the Union wishing to make [international] journeys are not free to object to the processing of their fingerprints. In those circumstances, persons applying for passports cannot be deemed to have consented to that processing.'
[40] See eg, Leenes and Kosta (n 13); Corien Prins and Lokke Moerel, 'Privacy for the Homo digitalis: Proposal for a new regulatory framework for data protection in the light of big data and the internet of things' (2016) <https://ssrn.com/abstract=2784123> accessed 7 September 2017.
[41] Kosta (n 34) 17.
[42] College Bescherming Persoonsgegevens, 'Brief aan de staatssecretaris van Onderwijs, Cultuur en Wetenschap, over beantwoording Kamervragen i.v.m. cookiebeleid [Letter to the State Secretary of Education, Culture and Science, on answers to parliamentary questions about cookie policy]' (31 January 2013) <https://autoriteitpersoonsgegevens.nl/sites/default/files/atoms/files/z2013-00718.pdf> accessed 7 September 2017; Natali Helberger, 'Freedom of Expression and the Dutch Cookie-Wall' (March

The Article 29 Working Party, in which European Data Protection Authorities cooperate, recommends not using tracking walls, but does not say that current law prohibits them. The Working Party says that people 'should have an opportunity to freely choose between the option to accept some or all cookies or to decline all or some cookies.'[43]

> In some Member States access to certain websites can be made conditional on acceptance of cookies, however generally, the user should retain the possibility to continue browsing the website without receiving cookies or by only receiving some of them, those consented to that are needed in relation to the purpose of provision of the website service, and those that are exempt from consent requirement. It is thus *recommended* to refrain from the use of consent mechanisms that only provide an option for the user to consent, but do not offer any choice regarding all or some cookies.[44]

Recital 25 of the ePrivacy Directive says that '[a]ccess to specific website content may still be made conditional on the well-informed acceptance of a cookie or similar device, if it is used for a legitimate purpose.' But the Working Party suggests that Recital 25 is not meant to allow companies to put the whole website behind a tracking wall: '[t]he emphasis on "specific website content" clarifies that websites should not make conditional "general access" to the site on acceptance of all cookies.'[45] The Working Party adds that website publishers should 'only limit certain content if the user does not consent to cookies.'[46] The careful phrasing suggests that the Working Party does not

---

Institute for Information Law, 2013) 18 <https://ssrn.com/abstract=2351204> accessed 7 September 2017.
[43] Article 29 Working Party 2013, 'Working Document 02/2013 providing guidance on obtaining consent for cookies' (2 October 2013) WP 208, 5.
[44] ibid 5 (internal footnote omitted, emphasis added).
[45] ibid 5.
[46] ibid 5.

mean to say that current law prohibits all tracking walls.[47] This seems to be the correct interpretation of current law. In the next section, we turn to future law: the GDPR.

## 2. The GDPR and Tracking Walls

Compared to the Data Protection Directive, the requirements for valid consent have been tightened in the GDPR. The GDPR defines consent as 'any freely given, specific, informed and unambiguous indication of the data subject's wishes by which he or she, by a statement or by a clear affirmative action, signifies agreement to the processing of personal data relating to him or her.'[48] Article 7 of the GDPR discusses when consent is freely given:

> When assessing whether consent is freely given, utmost account shall be taken of whether, *inter alia*, the performance of a contract, including the provision of a service, is conditional on consent to the processing of personal data that is not necessary for the performance of that contract.[49]

Hence, under Article 7 of the GDPR, to assess whether consent is freely given (and therefore valid), it must be considered whether a service is made conditional on consent. This rule can be applied to tracking walls. Web-wide tracking is not *necessary* for providing a website, according to the Article 29 Working Party.[50] The GDPR does not say that take-it-or-leave-it choices *always* lead to invalid consent. Rather, 'utmost

---

[47] But see the English Information Commissioner's Office, which says: 'the individual must have a genuine choice over whether or not to consent to marketing. Organisations should not coerce or unduly incentivise people to consent, or penalise anyone who refuses.' (Information Commissioner's Office, 'Direct Marketing. Data Protection Act, Privacy and Electronic Communications Regulations', Version 2.2, 14<http://ico.org.uk/for_organisations/guidance_index/~/media/documents/library/Privacy_and_electro nic/Practical_application/direct-marketing-guidance.pdf> accessed 7 September 2017).
[48] art 4(11) General Data Protection Regulation.
[49] art 7(4) General Data Protection Regulation. See generally on art 7(4) General Data Protection Regulation: Bojana Kostic and Emmanuel Vargas Penagos, 'The freely given consent and the "bundling" provision under the GDPR' (2017) 153 Computerrecht 217-222.
[50] Article 29 Working Party, 'Opinion 03/2013 on purpose limitation' (2 April 2013) WP 203, 46; Article 29 Working Party, 'Opinion 06/2014 on the notion of legitimate interests of the data controller under article 7 of Directive 95/46/EC' (9 April 2014) WP 217, 47.

account shall be taken' of whether a contract or service is made conditional on consent.[51]

The GDPR's recitals make the requirements for 'freely given' consent even stricter. Recitals in the preamble of a legal instrument of the EU have less legal weight than the instrument's provisions. Nevertheless, recitals can be used to interpret provisions.[52] Recital 43 of the GDPR states:

> In order to ensure that consent is freely given, consent should not provide a valid legal ground for the processing of personal data in a specific case where there is a clear imbalance between the data subject and the controller, in particular where the controller is a public authority and it is therefore unlikely that consent was freely given in all the circumstances of that specific situation. Consent is presumed not to be freely given if it does not allow separate consent to be given to different personal data processing operations despite it being appropriate in the individual case, or if the performance of a contract, including the provision of a service, is dependent on the consent despite such consent not being necessary for such performance.[53]

Recital 43 could be read as a presumption that tracking walls cannot be used to obtain valid consent. After all, a tracking wall typically makes providing a service dependent on the data subject's consent. Additionally, sometimes a data subject's consent may be invalid because of a clear imbalance between the data subject and a company. If there is a clear imbalance between a large company and a data subject, Recital 43 suggests that a tracking wall of such a company makes consent involuntary, and therefore invalid. When a large company, such as Facebook, processes personal data, it could be argued that a clear imbalance exists between the data subject and the company (the controller). After all, the data subject and Facebook do not have equal bargaining power; some Facebook users may feel that they have no choice to consent.

---

[51] art 7(4) General Data Protection Regulation.
[52] Todas Klimas and Jurate Vaičiukaitė, 'The Law of Recitals in European Community Legislation' (2008) 15(1) ILSA Journal of International & Comparative Law.
[53] recital 43 General Data Protection Regulation. The European Parliament had proposed a similar sentence in a provision (rather than in a recital); see, art 7(4) LIBE Compromise <http://www.janalbrecht.eu/fileadmin/material/Dokumente/DPR-Regulation-inofficial-consolidated-LIBE.pdf> accessed 7 September 2017.

The GDPR's preamble gives more arguments for invalidating consent when a website uses a tracking wall. Recital 42 says: 'Consent should not be regarded as freely given if the data subject has no genuine or free choice or is unable to refuse or withdraw consent without detriment.'[54] If not being able to use a website is seen as 'detriment', this recital seems to make consent invalid when a website uses a tracking wall. Recital 32 adds that '[i]f the data subject's consent is to be given following a request by electronic means, (…) the request must be (…) not unnecessarily disruptive to the use of the service for which it is provided.' As far as tracking walls are unnecessarily disruptive, they do not comply with this recital.

In conclusion: the GDPR does not categorically prohibit tracking walls and similar take-it-or-leave-it choices. But compared to the Data Protection Directive, the GDPR is much stricter on the conditions under which consent is 'freely given'.

## VI. The Proposed ePrivacy Regulation

In January 2017, the European Commission published a proposal for an ePrivacy Regulation.[55] Like the ePrivacy Directive, the ePrivacy proposal contains a provision that applies to cookies and online tracking. Article 8(1) of the ePrivacy proposal reads as follows:

---

[54] recital 41 General Data Protection Regulation.
[55] Proposal for a Regulation of the European Parliament and of the Council, concerning the respect for private life and the protection of personal data in electronic communications and repealing Directive 2002/58/EC (Regulation on Privacy and Electronic Communications) COM(2017) 10 final <https://ec.europa.eu/digital-single-market/en/news/proposal-regulation-privacy-and-electronic-communications> accessed 7 September 2017.

The use of processing and storage capabilities of terminal equipment and the collection of information from end-users' terminal equipment, including about its software and hardware, other than by the end-user concerned shall be prohibited, except on the following grounds:

(a) it is necessary for the sole purpose of carrying out the transmission of an electronic communication over an electronic communications network; or

(b) the end-user has given his or her consent; or

(c) it is necessary for providing an information society service requested by the end-user; or

(d) if it is necessary for web audience measuring, provided that such measurement is carried out by the provider of the information society service requested by the end-user.[56]

Roughly speaking, Article 8(1) of the ePrivacy proposal replaces Article 5(3) of the current ePrivacy Directive. The proposed Article 8(1) applies to cookies and to many other tracking technologies; a cookie 'use[s] processing and storage capabilities of terminal equipment'.[57] Like the ePrivacy Directive, the ePrivacy proposal refers to general data protection law (the GDPR) for the definition of consent.[58] The ePrivacy proposal does not contain specific rules on tracking walls.[59]

## VII.  Four Options to Regulate Tracking Walls

How could the lawmaker deal with tracking walls? We discuss four options: (i) no specific rules for tracking walls; (ii) ban tracking walls in certain circumstances; (iii) fully ban tracking walls; (iv) ban all web-wide tracking. (A general discussion of Article 8(1) of the ePrivacy proposal falls outside the scope of this paper; we focus on tracking walls.[60])

---

[56] art 8(1) ePrivacy proposal. See also recitals 6, 20, and 21 of the ePrivacy proposal.
[57] art 8(1) ePrivacy proposal.
[58] art 9(1) ePrivacy proposal.
[59] The ePrivacy proposal does mention take-it-or-leave-it choices in the context of art 6, in recital 18.

## 1. Option (i): No Specific Rules for Tracking Walls

A first option for the lawmaker is: not including specific rules on tracking walls in the ePrivacy Regulation. If the lawmaker does not add specific rules, the voluntariness of consent would have to be assessed separately for each tracking wall. The main question would be, in each case, whether people can 'freely' give consent in line with the GDPR's requirements.

Below we provide a first sketch for a circumstance catalogue, a non-exhaustive list of circumstances in which it is particularly questionable whether consent is valid if a company offers a take-it-or-leave-it choice, for instance with a tracking wall.

If a company has a dominant position, there is more chance of imbalance between the contract parties as individuals have little negotiation power vis-à-vis such a company. As Bygrave notes, 'fairness implies a certain protection from abuse by data controllers of their monopoly position.'[61]

Sometimes, for other reasons it is not a realistic option for people to go to a competitor, for instance because of a lock-in situation. In a lock-in situation, it is difficult or costly to leave a service.[62] To illustrate, when all one's friends are on Facebook, joining another social network site makes little sense. In such a case, it is hardly a realistic option to go to a privacy-friendly competitor.

It is dubious whether consent is still 'freely given' if a company uses a tracking wall and there are no competitors that offer a similar, more privacy-friendly service.[63]

---

[60] See generally on the ePrivacy proposal, Zuiderveen Borgesius (n 3).
[61] Lee Bygrave, 'Data protection law: approaching its rationale, logic and limits' (PhD thesis, University of Oslo) (Kluwer Law International 2002) 58.
[62] Hal Shapiro and Carl Varian, *Information Rules. A Strategic Guide to the Network Economy* (Harvard Business School Press 1999) 104. In a lock-in situation, 'the costs of switching from one brand of technology to another are substantial.'
[63] See s 28(3)(b) Federal Data Protection Act in Germany.

Somebody who does not want to disclose personal data would not have the possibility to use a certain type of service.[64]

Additionally, some people may be more vulnerable to pressure – such situations call for a more privacy-protective interpretation of the rules. To illustrate: a tracking wall is questionable when a service is aimed at, or often used by, children.[65] After all, children are less likely to fully understand the implications of that choice, or might feel more readily pressured into accepting what they may perceive as a non-choice. Similarly, people with a medical condition might be more easily pressured into consenting to data collection if they believe that access to a particular website will give them important health information. In conclusion, all circumstances should be taken into account to assess the voluntariness, and thus the validity, of consent.

But the GDPR's requirements regarding consent remain open for conflicting interpretations. More guidance on the legality of tracking walls could improve legal clarity. Case law could clarify the conditions under which consent should be considered to be 'freely given'. But it may take a long time until there is enough case law to have clarity. Therefore, specific rules on tracking walls might be preferable.

## 2. Option (ii) Ban Tracking Walls in Certain Circumstances

A second option is banning tracking walls under certain circumstances. Indeed, in some circumstances, tracking walls should not be allowed at all. Or, to be more exact: in some circumstances, the law should not allow companies to use a tracking wall to obtain valid consent for tracking.

---

[64] Especially in markets with information asymmetry, there might not be any privacy-friendly competitors. See, Tony Vila, Rachael Greenstadt and David Molnar, 'Why We Can't be Bothered to Read Privacy Policies. Models of Privacy Economics as a Lemons Market' in Jean Camp and Stephen Lewis (eds), *Economics of Information Security* (Springer 2004); Frederik J Zuiderveen Borgesius, 'Behavioural Sciences and the Regulation of Privacy on the Internet' in Anne Sibony and Alberto Alemanno (eds), "*Nudging and the Law - What can EU Law learn from Behavioural Sciences?* (Hart Publishing).

[65] See, Article 29 Working Party: 'The user should be put in a position to give free and specific consent to receiving behavioural advertising, independently of his access to the social network service.' Perhaps this remark is partly inspired by the fact that many children use social network sites [Article 29 Working Party, 'Opinion 15/2011 on the definition of consent' (n 37) 18].

For instance, in 2016, the Article 29 Working Party mentioned five circumstances in which tracking walls should be banned. In short, the Working Party called for a prohibition of tracking walls: (i) on websites that reveal special categories of data (sensitive data) to trackers; (ii) in the case of 'tracking by unidentified third parties for unspecified purposes';[66] (iii) on government-funded websites; (iv) in circumstances where the GDPR would make consent invalid; (v) in the case of 'bundled consent for processing for multiple purposes. Consent should be granular.'[67] (In 2017, the Working Party amended its position; now it calls for a complete ban of tracking walls.[68])

The five circumstances mentioned by the Working Party are a good start. But tracking walls should probably be banned in more circumstances. So far, we have assessed tracking walls primarily from the perspective of data protection law and the ePrivacy Directive. But considerations from other policy areas can help to inform the debate about tracking walls.

Our survey shows that most people think tracking walls are unfair, regardless of the type of website. However, there are arguments based on public policy considerations in favour of introducing a distinction between different websites, or types of websites, when considering rules for tracking walls.

Below, we discuss public policy arguments for specific rules or bans in certain circumstances, in particular arguments relating to freedom of expression and media, professions with confidentiality requirements, and the public sector. We do not mean to give an exhaustive list; we mention these examples as a starting point for a discussion.

---

[66] Here, the Working Party probably refers to 'real time bidding'. See Lukasz Olejnik, Tran Minh-Dung and Claude Castelluccia, 'Selling Off Privacy at Auction' (2013) Inria <https://hal.inria.fr/hal-00915249> accessed 7 September 2017.
[67] Article 29 Working Party, 'Opinion 03/2016 on the evaluation and review of the ePrivacy Directive (2002/58/EC)' (26 July 2016) WP 240, 17.
[68] Article 29 Working Party, 'Opinion 01/2017 on the Proposed Regulation for the ePrivacy Regulation (2002/58/EC)' (n 7).

## a. Freedom of Expression

One relevant area is media law. In media law, the right to freedom of expression is a guiding principle. The right to freedom of expression in the European Convention on Human Rights (Article 10) and in the Charter of Fundamental Rights (Article 11) includes the freedom to *receive* information and ideas without interference by public authority.[69] And the International Covenant on Civil and Political Rights protects the right to 'seek' information.[70]

Article 10 of the European Convention on Human Rights does not merely require states to refrain from interfering with the right to freedom of expression. Sometimes, states must take action to safeguard freedom of expression, says the European Court of Human Rights: 'the genuine and effective exercise of freedom of expression under Article 10 may require positive measures of protection, even in the sphere of relations between individuals.'[71] Article 10 does not grant a right to receive media content free of charge, or a right not to be tracked online. Nevertheless, Article 10 illustrates the importance of the public's interest in receiving information.

Web-wide tracking may threaten the public's freedom to receive information. If access to media content is made conditional (upon conditions such as the obligation to pay a price or to accept tracking cookies), parts of the public may be excluded. For instance, privacy-aware people may not want to accept tracking cookies and may therefore not access content that is behind tracking walls. Moreover, tracking people's online behaviour could lead to a chilling effect. For example, people might refrain from visiting certain websites, or from reading about certain political topics, if they fear that their behaviour is tracked. People may want to keep their political views confidential, and may not want others (including tracking companies) drawing the wrong conclusions about their political opinions. Indeed, data protection law recognises personal data regarding political opinions as data that deserve extra protection.[72]

---

[69] art 10 European Convention on Human Rights. The European Court of Human Rights (ECtHR) says that 'the public has a right to receive information of general interest' [*Társaság a Szabadságjogokért v Hungary* App no 37374/05 (ECtHR, 14 April 2009) para 26]. And 'the internet plays an important role in enhancing the public's access to news and facilitating the sharing and dissemination of information generally (…).' [*Fredrik Neij and Peter Sunde Kolmisoppi (The Pirate Bay) v Sweden* App no 40397/12 (ECtHR, 19 February 2013 (inadmissible))].

[70] art19(2) International Covenant on Civil and Political Rights. See generally on the right to receive information, Sarah Eskens, 'Challenged by news personalization: Five perspectives on the right to receive information' (forthcoming).

[71] *Khurshid Mustafa v Sweden* App no 23883/06 (ECtHR, 16 March 2009) para 32.

Hence, privacy can be important for the right to receive information. As Richards notes, 'a specific kind of privacy is *necessary* to protect our cherished civil liberties of free speech and thought.'[73] He speaks of 'intellectual privacy', the 'protection from surveillance or interference when we are engaged in the processes of generating ideas – thinking, reading, and speaking with confidants before our ideas are ready for public consumption.'[74] Along similar lines, Cohen argues for a 'right to read anonymously.'[75]

Chilling effects are hard to prove empirically. Nevertheless, there is some evidence that surveillance by the state[76] and by companies[77] causes a chilling effect. More empirical research is needed into chilling effects resulting from commercial tracking and surveillance. If a chilling effect occurred in relation to reading about news and politics, it could threaten our democratic society. This could be an argument to consider bans on tracking walls on media websites.

b. Institutions with a Public Mission: Media

Specific rules should be considered for public service media.[78] Tracking walls for public service media can conflict with the mission of the public service media.

---

[72] art 9 GDPR; art 8 Data Protection Directive.
[73] Neil Richards, *Intellectual Privacy: Rethinking Civil Liberties in the Digital Age* (Oxford University Press 2014) 9.
[74] ibid 9. See also, JE Cohen, 'Intellectual Privacy and Censorship of the Internet' (1997) 8 Seton Hall Constitutional Law Journal 693 and Declaration of the Committee of Ministers on Risks to Fundamental Rights stemming from Digital Tracking and other Surveillance Technologies (Adopted by the Committee of Ministers on 11 June 2013) para 2; Joined Cases C-203/15 and C-698/15 *Tele2 Sverige AB and Watson* [2016] ECLI:EU:C:2016:970, para 101.
[75] JE Cohen, 'A Right to Read Anonymously: A Closer Look at Copyright Management in Cyberspace' (1995) 28 Connecticut Law Review 981.
[76] See, Jon Penney, 'Chilling Effects: Online Surveillance and Wikipedia Use' (2016) 31(1) Berkeley Technology Law Journal, art 5; Alex Marthews and Catherine Tucker, 'Government Surveillance and Internet Search Behavior' (29 April 2015) <http://ssrn.com/abstract=2412564> accessed 7 September 2017.
[77] A survey by Cranor and McDonald in 2010 suggests behavioural targeting has a chilling effect, but the research concerns declared (not revealed) preferences. McDonald and Cranor (n 19)..
[78] As noted, the Dutch Data Protection Authority criticised the public broadcaster for using a tracking wall. Dutch Data Protection Authority (*Autoriteit Persoonsgegevens*), 'Beantwoording Kamervragen

According to the Council of Europe, public service media should promote democratic values, and should offer 'universal access.'[79] A situation in which the more privacy-conscious people are excluded from access to the programs of the public service media could conflict with the requirement of universal access.

Should specific rules be adopted for tracking walls on news websites and other commercial media? That is an even harder question than the question regarding public service media. The mission of commercial media is harder to define than the mission of public service media. And many commercial publishers, more than public service media, depend on advertising income. Moreover, commercial media can invoke their 'freedom to conduct a business in accordance with Union law and national laws and practices', which is granted in the Charter of Fundamental Rights of the European Union.[80] Commercial website publishers could argue that the right to conduct a business implies that they can set the conditions for access.[81]

A ban of tracking walls on news websites could lower behavioural advertising income for news publishers, at least in the short term. However, as noted, in the long term, behavioural advertising may diminish profit for news websites.[82] And even if tracking-based or behavioural advertising were not allowed, other types of online advertising, such as contextual advertising, would still be possible.

Nevertheless, lawmakers should be cautious with rules that reduce income for news websites. A ban on tracking walls for news websites would require further debate and research. For instance, it would be hard to define the scope of the ban. Should the ban apply to political blogs, or to online newspapers that only gossip about celebrities?

---

i.v.m. cookiebeleid NPO' (31 January 2013) 5 <https://autoriteitpersoonsgegevens.nl/sites/default/files/atoms/files/med_20130205-cookies-npo.pdf> accessed 7 September 2017.

[79] Recommendation CM/Rec(2007)3 of the Committee of Ministers to member states on the remit of public service media in the information society (31 January 2007) para 1(a). See also, art 2(f) and 3 Dutch media law.

[80] art 16 Charter of Fundamental Rights of the European Union.

[81] See Interactive Advertising Bureau Europe, 'Position on the proposal for an ePrivacy Regulation' (Position paper, 28 March 2017) <https://iabireland.ie/wp-content/uploads/2017/03/20170328-IABEU-ePR_Position_Paper.pdf> accessed 7 September 2017.

[82] See s II.

Lawmakers could also take less drastic measures than banning tracking walls. For instance, lawmakers could require publishers to offer a tracking-free version of their website, which visitors must pay for with money.[83] (As noted, online advertising is possible without tracking and data collection. Hence, a tracking-free version could still contain advertising.)

However, as the European Data Protection Supervisor notes, 'privacy is not a luxury but a universal right and it should not only be available to those with the means to pay.'[84] And many people say it is 'extortion' if they have to pay for privacy.[85] In a EU-wide survey, 74% finds 'paying not to be monitored when using a website' unacceptable.[86] Hence, a legal requirement for a tracking-free version of websites that people must pay for with money has drawbacks. Nevertheless, such a requirement could give many people more choice: now a tracking-free version is often not available at all.

Media law could also incorporate the conditions that need to be fulfilled to safeguard media users' interests and broader societal objectives. Privacy rules in media law would not be a complete novelty. Under the German Telemedia Act, a

> service provider must enable the use of telemedia and payment (…) to occur anonymously or via a pseudonym where this is technically possible and reasonable. The recipient of the service is to be informed about this possibility.[87]

In sum, media policy arguments can be taken into account when regulating tracking walls.

---

[83] Peter Traung, 'The Proposed New EU General Data Protection Regulation: Further Opportunities' (2012)(2) Computer Law Review International 33, 42; Helberger (n 41). See also, European Commission, 'Summary report on the public consultation on the Evaluation and Review of the ePrivacy Directive' (Consultation Results, 4 August 2016) Question 22 <https://ec.europa.eu/digital-single-market/en/news/summary-report-public-consultation-evaluation-and-review-eprivacy-directive> accessed 7 September 2017.

[84] European Data Protection Supervisor, 'EDPS Opinion on coherent enforcement of fundamental rights in the age of big data' (23 September 2016) 16 https://edps.europa.eu/sites/edp/files/publication/16-09-23_bigdata_opinion_en.pdf accessed 28 September 2017 .

[85] See McDonald and Cranor (n 19). 27.

[86] Commission. 'Flash Eurobarometer 443' (n 22) Question 5.3 (95/T.26).

[87] art 13(4) No 6 of the German Telemedia Law (*Telemediendienste-Gesetz*).

## c. Professional Secrecy

Specific rules could also be considered for parties with professional secrecy requirements, such as doctors, lawyers, and accountants. For instance, hospital websites should not use tracking walls. A patient should be able to make an appointment with a cancer specialist through a hospital website, without fearing that marketing companies see that the patient does so. In the medical sector, there is a long tradition of protecting personal data regarding health, as illustrated by the Hippocratic oath, which requires doctors to keep patient information confidential. Medical secrecy protects individual privacy interests of patients, and a public interest: the trust in medical services.[88] As the European Court of Human Rights notes:

> The protection of personal data, in particular medical data, is of fundamental importance to a person's enjoyment of his or her right to respect for private and family life as guaranteed by Article 8 of the Convention. Respecting the confidentiality of health data is a vital principle in the legal systems of all the Contracting Parties to the Convention. It is crucial not only to respect the sense of privacy of a patient but also to preserve his or her confidence in the medical profession and in the health services in general.[89]

At a minimum, hospital websites should give people the chance to visit a tracking-free version of the website.[90] As noted, the Article 29 Working Party proposed a ban of tracking walls on sites that expose visitors' special categories of data (sensitive data) to tracking companies.[91] Such a ban would also lead to the conclusion that hospitals cannot use tracking walls. In sum, a ban on tracking walls seems appropriate for hospital websites. The European Consumers' Organisation goes further, and says: 'Tracking and profiling technologies in health-related web sites should not be allowed.'[92]

---

[88] Martine Ploem, 'Tussen Privacy en Wetenschapsvrijheid. Regulering van Gegevensverwerking voor Medisch-Wetenschappelijk Onderzoek' [Between Privacy and Scientific Freedom. Regulation of Data Processing for Medical Scientific Research] (PhD thesis, University of Amsterdam) (SDU 2004) 129-133.

[89] *I v Finland* App no 25011/03 (ECtHR,17 July 2008) para 38. See along similar lines, *Z v Finland* App no 22009/93 (ECtHR, 25 February 1997) para 95 and European Union Civil Service Tribunal, Civil Service Tribunal Decision F-46/095, *V & EDPS v European Parliament* (5 July 2011) paras 123 and 163.

[90] See, Article 29 Working Party, 'Opinion 03/2016 on the evaluation and review of the ePrivacy Directive (2002/58/EC)' (n 66) 17; Bart Jacobs, 'Klantverraders' [Customer snitches] (*PI.lab*, 13 December 2015) <http://pilab.nl/about%20pi%20lab/blog/klantverraders.html> accessed 7 September 2017.

It would be hard to phrase such prohibitions in a way that does not make them over or under inclusive. How to define 'health related' websites? Is it enough if the website presents itself as a health-related website, for instance by including a picture of a doctor in a white coat? And would a prohibition of using any 'health data' for behavioural targeting also cover tracking of daily visits to a website with gluten free recipes?

### d. Public Sector

Specific rules could also be considered for the public sector. There are precedents: in the Netherlands, tracking walls on public sector websites are banned.[93] Indeed, people should be able to visit public sector websites without exposing themselves to web-wide tracking. In practice, public sector websites might use third party widgets such as social media buttons.[94] The public sector body might not realise that it exposes visitors to web-wide tracking when it includes such widgets on its website.

More generally, it is debatable whether it is appropriate for public sector bodies to allow web-wide tracking for commercial purposes on their websites – even if people consent. The lawmaker could consider banning all web-wide tracking on public sector websites. The exact scope of such a ban would require further debate. For instance, should the ban apply to organisations that are partly funded by the state? And some site-wide tracking could be necessary for website security,[95] or useful for website analytics.[96]

---

[91] Article 29 Working Party, 'Opinion 03/2016 on the evaluation and review of the ePrivacy Directive (2002/58/EC)' (n 66) 17.

[92] European Consumer Organisation BEUC, 'Data Protection Regulation. Proposal for a Regulation' (BEUC Position paper, August 2012) 8 <http://www.beuc.eu/publications/2012-00531-01-e.pdf> accessed 7 September 2017.

[93] art 11.7a(5) Telecommunications Act (*Telecommunicatiewet*). See in English, Eleni Kosta, 'The Dutch regulation of cookies' (2016) 2(1) European Data Protection Law Review 97.

[94] To illustrate: Van Der Velden found third party tracking on many Dutch governmental websites. Lonneke van der Velden, 'The third party diary: tracking the trackers on Dutch governmental websites' (2014) 3(1) NECSUS European Journal of Media Studies 195.

## e. A Black List and a Grey List

To sum up: bans should be considered for tracking walls, at least in certain circumstances. As a starting point for a discussion, we discussed public service media, commercial media, professions with specific confidentiality rules, and the public sector. A black list of circumstances in which tracking walls are banned, should be non-exhaustive. In other words: depending on the circumstances, a tracking wall can make consent involuntary (and thus invalid), even if the situation is not explicitly included on the black list.

The lawmaker could consider supplementing the black list (with prohibitions) with a grey list. If a situation is on the grey list, there is a legal presumption that a tracking wall makes consent involuntary, and therefore invalid. Hence, the legal presumption of the grey list shifts the burden of proof. For situations on the grey list, it is up to the company employing the tracking wall to prove that people can give 'freely given' consent, even though the company installed a tracking wall. Some consumer protection laws use a similar system, with a black list (illegal practices) and a grey list (practices presumed to be illegal).[97] The situations we discussed in our circumstance catalogue, with circumstances in which the voluntariness of consent is questionable, could be included on the grey list.

## 3. Option (iii): Ban Tracking Walls Completely

A third option is banning tracking walls completely. Both the Article 29 Working Party and the European Data Protection Supervisor call for a complete ban.[98] The European

---

Parliament's rapporteur for the ePrivacy Regulation, Lauristin,[99] and civil society organisations also argue for a complete ban on tracking walls.[100] It appears that a complete ban on tracking walls would be popular among the general public. Our survey shows that many people think tracking walls are unfair and unacceptable.[101]

If tracking walls were banned, a website publisher could still ask visitors whether they want to be tracked for targeted advertising. The main effect of the ban would be that the publisher could not offer people a take-it-or-leave-it choice regarding web-wide tracking (third party tracking). We emphasise that a ban on tracking walls would not interfere with cookies that can be set without consent (for instance for a service requested by the end-user).[102] Generally speaking, the ban on tracking walls should probably only apply to web-wide tracking.

## 4. Option (iv): Ban All Web-Wide Tracking

A fourth option is more extreme: ban all web-wide tracking (third party tracking). Some have argued for such a prohibition. For instance, according to security technologist Schneier it is 'vital' to adopt such a 'ban on third party ad tracking'. He adds: 'it's the companies that spy on us from website to website, or from device to device, that are

---

24 April 2017 <https://edps.europa.eu/sites/edp/files/publication/17-04-24_eprivacy_en.pdf> accessed 7 September 2017, p. 17.

[99] Committee on Civil Liberties, Justice and Home Affairs, Draft Report on the proposal for a regulation of the European Parliament and of the Council concerning the respect for private life and the protection of personal data in electronic communications and repealing Directive 2002/58/EC (Regulation on Privacy and Electronic Communications) (COM(2017)0010 – C8-0009/2017 – 2017/0003(COD)), 9 June 2017 <http://www.europarl.europa.eu/sides/getDoc.do?pubRef=-%2F%2FEP%2F%2FNONSGML%2BCOMPARL%2BPE-606.011%2B01%2BDOC%2BPDF%2BV0%2F%2FEN> accessed 7 September 2017.

[100] European Digital Rights, 'EDRi's position on the proposal of an e-privacy regulation', April 2017 <https://edri.org/files/epd-revision/ePR_EDRi_position_20170309.pdf> accessed 7 September 2017; La Quadrature du Net, 'Recommendations of La Quadrature du Net on the revision of the eprivacy Directive' 6 March 2017 <https://www.laquadrature.net/files/ePrivacy_LQDN_recommendations_060317.pdf> accessed 5 May 2017.

[101] See Section IV.

[102] See the exceptions in Article 5(3) of the ePrivacy Directive and Article 8(1) of the ePrivacy proposal.

doing the most damage to our privacy.'[103] Along similar lines, Ceglowski argues for these rules:

> Sites showing ads may only use two criteria in ad targeting: the
> content of the page, and whatever information they have about
> the visitor. Sites may continue to use third-party ad networks to
> serve ads, but those third parties must be forgetful; they may
> not store any user data across ad requests.[104]

Such a ban on web-wide tracking would probably improve the protection of privacy and personal data. An additional advantage of a complete ban is its simplicity and legal clarity. However, we do not discuss the option of banning all web-wide tracking in detail here. A complete ban on web-wide tracking has drawbacks. For example, some people might prefer targeted (tracking-based) ads to non-targeted ads.[105] And a ban on web-wide tracking would interfere with many business models. We think it makes sense to start with lighter measures than a complete ban on web-wide tracking. If such lighter measures do not lead to appropriate protection of privacy and other fundamental rights, a complete ban could be considered.

In conclusion, we think that the lawmaker should seriously consider option (ii), a partial ban of tracking walls, and option (iii), a complete ban on tracking walls. An advantage of option (ii) is that banning tracking walls only under certain circumstances is a more nuanced approach than completely banning them. The partial ban (the black list) could be complemented with a grey list (circumstances in which tracking walls are presumed to be illegal). A disadvantage of option (ii) is that the nuance comes at the cost of legal clarity. A major advantage of option (iii), a complete ban of tracking walls, is that such a rule could be phrased in a relatively clear and straightforward way. Hence, a complete ban would provide more legal clarity than a partial ban.

---

[103] Schneier, 'The Internet of Things That Talk About You Behind Your Back' (8 January 2016) <https://motherboard.vice.com/en_us/article/the-internet-of-things-that-talk-about-you-behind-your-back> accessed on 31 August 2017.
[104] Ceglowski, 'What Happens Next Will Amaze You', Talk at Fremtidens Internet Conference in Copenhagen, Denmark (14 September 2015) <http://idlewords.com/talks/what_happens_next_will_amaze_you.htm> accessed 7 September 2017.
[105] See: Ur et al., 'Smart, Useful, Scary, Creepy: Perceptions of Online Behavioral Advertising' (Proceedings of the Eighth Symposium on Usable Privacy and Security ACM, 2012) 4.

## VIII.  Take-It-Or-Leave-It Choices in Other Contexts

We focused our discussion on one specific type of take-it-or-leave-it choices: tracking walls. But take-it-or-leave-it choices regarding privacy also occur in other situations. For instance, chat and email services often require users to agree to a data use policy – if people do not agree, they cannot use the service. An app might require access to the camera or the contact list on an end-user's phone, while that access is unnecessary for providing the service. A 'smart' TV might listen to the sounds in people's living rooms, and might only work when people 'consent' to that. Internet of Things equipment may only work if people 'consent' to data collection for marketing.

We call for a more general debate on the appropriateness of such take-it-or-leave-it choices. The EU lawmaker should consider adopting additional rules regarding take-it-or-leave-it choices regarding privacy that do not concern tracking on websites.[106] In the shorter term, the Article 29 Working Party or the European Data Protection Board could adopt guidance on the interpretation of the GDPR's requirements for valid and 'freely given' consent.[107]

## IX.  Conclusion

On the internet, many companies offer people take-it-or-leave-it-choices regarding privacy. For instance, some websites install tracking walls, barriers that visitors can only pass if they click 'OK' to a request to place tracking cookies to monitor their online behaviour. Our survey in the Netherlands shows that most people think tracking walls are unfair and unacceptable. But if people encounter such tracking walls, they are likely to consent, even if they think it is not fair that they have to disclose personal data in exchange for using a website or other service.

---

[106] The EDPS suggests a specific provision that bans take-it-or-leave-it choices regarding privacy and personal data in the context of the Internet of Things. European Data Protection Supervisor (n 97) 18.
[107] The Article 29 Working Party plans to publish guidance on 'consent' in 2017: 'Adoption of 2017 GDPR action plan' (16 January 2017) <http://ec.europa.eu/newsroom/document.cfm?doc_id=41387> accessed 7 September 2017.

European data privacy law gives a central role to informed consent of the internet user. For valid consent, the GDPR requires a 'freely given' indication of wishes. The exact meaning of 'freely given' is unclear. We suggested a list of circumstances to consider when assessing whether a tracking wall is allowed under the GDPR. We also gave starting points for a broader debate: perhaps more legal intervention is needed. A partial or complete ban of tracking walls should be considered. More generally, research and debate is needed on take-it-or-leave-it choices regarding privacy.

* * *